\documentclass[prd,preprint,tightenlines,showpacs,preprintnumbers,nofootinbib,eqsecnum,superscriptaddress]{revtex4}

\usepackage{dcolumn}
\usepackage{bm}
\usepackage{epsfig}
\usepackage{amsfonts}
\usepackage{amsmath}
\usepackage{amssymb,amscd}

\def\lsim{\raise0.3ex\hbox{$<$\kern-0.75em\raise-1.1ex\hbox{$\sim$}}}

\def\gsim{\raise0.3ex\hbox{$>$\kern-0.75em\raise-1.1ex\hbox{$\sim$}}}

\newcommand{\be}{\begin{equation}}

\newcommand{\ee}{\end{equation}}

\def\beq{\begin{equation}}

\def\eeq{\end{equation}}

\def\beqa{\begin{eqnarray}}

\def\eeqa{\end{eqnarray}}

\newcommand{\ba}{\begin{eqnarray}}

\newcommand{\ea}{\end{eqnarray}}

\def\gappeq{\mathrel{\rlap {\raise.5ex\hbox{$>$}}

{\lower.5ex\hbox{$\sim$}}}}

\def\lappeq{\mathrel{\rlap{\raise.5ex\hbox{$<$}}

{\lower.5ex\hbox{$\sim$}}}}

\def\Toprel#1\over#2{\mathrel{\mathop{#2}\limits^{#1}}}

\begin{document}

\title{Muon trident process at far-forward LHC detectors}

\author{Reinaldo {\sc Francener}}
\email{reinaldofrancener@gmail.com}
\affiliation{Instituto de Física Gleb Wataghin - Universidade Estadual de Campinas (UNICAMP), \\ 13083-859, Campinas, SP, Brazil. }

\author{Victor P. {\sc Gon\c{c}alves}}
\email{barros@ufpel.edu.br}
\affiliation{Institute of Physics and Mathematics, Federal University of Pelotas, \\
  Postal Code 354,  96010-900, Pelotas, RS, Brazil.}

\author{Gabriel {\sc Rabelo-Soares}}
\email{rabelosoares21@gmail.com}
\affiliation{Instituto de Física Gleb Wataghin - Universidade Estadual de Campinas (UNICAMP), \\ 13083-859, Campinas, SP, Brazil. }

\begin{abstract}
The electromagnetic production of a dilepton pair in the muon-ion scattering, usually denoted muon trident process, is investigated considering the feasibility of studying processes induced by muons at LHC using its far-forward detectors. The total and differential cross-sections are estimated taking into account of  the Bethe-Heitler and bremsstrahlung channels, and predictions for the event rates expected for muon-tungsten ($\mu W$) collisions at the LHC energies are presented. Our results indicate that the observation of the muon trident process is feasible at FASER$\nu$. In particular, our results indicate  that the electromagnetic production of $\tau^+ \tau^-$ pairs in the $\mu W$ scattering    can, in principle,  be observed for the first time ever at the LHC. In addition, we present our predictions for the production of QED bound states. Finally, results for the trident process at the FASER$\nu$2 detector are also presented.
\end{abstract}



\maketitle

\vspace{1cm}
\section{Introduction}
Over the last decades, several theoretical \cite{Bjorken:1966kh,Tannenbaum:1968zz,Albright:1977zn, Albright:1977rk, Albright:1978mg,Ganapathi:1978qm,Ganapathi:1979mn, Ganapathi:1979dj,Barger:1979eg,Ganapathi:1980wh, Kelner:1998mh,Altmannshofer:2019zhy, Ballett:2018uuc,Zhou:2019vxt,beacom2, Francener:2024wul, Altmannshofer:2024hqd, Abbiendi:2024swt,Lee:2023cwb, Bigaran:2025vea, Francener:2024jra} and experimental \cite{dielectrons, Russell:1971mf,CHARM-II:1990dvf,CCFR:1991lpl,NuTeV:1999wlw,Maciuc:2006xb,SNDLHC:2024bzp} groups have performed the study of the trident process, i.e., the trilepton production by the scattering of lepton beams in a hadronic target, motivated by the possibility of testing the Standard Model and  searching for signals of New Physics.
The interest in this process has been recently renewed with the discovery of collider neutrinos by the FASER \cite{FASER:2023zcr,FASER:2024hoe,FASER:2024ref} and SND@LHC \cite{SNDLHC:2023pun} collaborations, which have verified that an intense and collimated flux of neutrinos is produced in the forward direction of $pp$ collisions at the LHC, as predicted in Ref. \cite{DeRujula:1984pg}.  In particular, Refs. \cite{Francener:2024wul,Altmannshofer:2024hqd}    have estimated the neutrino trident production, which is a weak process characterized by the production of a pair of charged leptons through the neutrino scattering in the Coulombian field of a heavy nucleus, considering neutrino-tungsten scattering  at the far-forward LHC detectors, and demonstrated that  a future observation of this rare process is feasible.

In addition to the neutrino flux, $pp$ collisions at the LHC also generate a flux of muons~\cite{DeRujula:1984pg} that are able to reach the far-forward LHC detectors. The measurement of such flux by the SND@LHC detector \cite{SNDLHC:2023mib} has started a new era in the study of  deep inelastic muon-ion scattering \cite{Francener:2025pnr,Francener:2025tyh}, which will allow us to improve the description of the hadronic structure and constrain the magnitude of nuclear effects in a new kinematical range.
In addition, several recent works have explored the potential of this muon flux to search for beyond Standard Model signals \cite{Ariga:2023fjg,Batell:2024cdl,MammenAbraham:2025gai}.
Motivated by these results, in this paper, we will investigate the muon trident process, which is electromagnetic production of a dilepton pair in the muon-ion scattering. In particular, we will consider the production of $e^+ e^-$, 
$\mu^+ \mu^-$ and $\tau^+ \tau^-$ pairs in the muon-tungsten scattering at the existing FASER$\nu$ detector and at its proposed upgrade to be installed in the Forward Physics Facility \cite{Anchordoqui:2021ghd,Feng:2022inv,FPF:2025bor}, denoted FASER$\nu$2 detector. As we will demonstrate below, the predicted event rates for the $e^+ e^-$ and $\mu^+ \mu^-$ production are very high, especially for the FASER$\nu$2 detector, which will allow us to constrain the magnitude of the associated cross-sections that are important, e.g., to estimate the electromagnetic energy loss of a muon in transit through materials \cite{Bulmahn:2008fa}. Moreover, our results indicate that a future measurement of the electromagnetic $\tau^+ \tau^-$ production will already be, in principle, feasible at FASER$\nu$. It is important to emphasize that such a process was not yet observed.

Motivated by the results for the lepton pair production in muon-tungsten interactions, we will also investigate the possibility that the produced dileptons create a QED bound state. In  recent years, the formation of a true muonium, i.e., a muon-antimuon bound state, by photon-photon interactions has been estimated considering  hadron-hadron \cite{Ginzburg:1998df,Azevedo:2019hqp,Francener:2021wzx,Dai:2024imb,Bertulani:2023nch,Gninenko:2025hsv,Liang:2025sol}, electron-hadron \cite{Francener:2024eep}, electron-positron \cite{Gargiulo:2023tci,Gargiulo:2024zyc,Gargiulo:2025pmu} collisions. 
Despite the effort, this bound state has never been observed. In this paper, we will  complement these previous studies by investigating the production of the true muonium in muon-tungsten interactions at the far-forward LHC detectors. For completeness, results for the positronium production will also be presented.

This paper is organized as follows. In the next section, we present a brief review of the formalism and tools used in our calculations. In Section \ref{sec:res}, we present our results for the energy dependence of the total cross-sections and differential distributions considering muon-tungsten interactions at FASER$\nu$. Moreover, the energy dependence of the cross-section for the true muonium is also discussed. Results for the event rates at FASER$\nu$ and FASER$\nu$2 are presented. Finally, in Section \ref{sec:sum}, we summarize our main results and conclusions.

\begin{figure}[t]
	\centering
	\begin{tabular}{ccc}
    \includegraphics[scale=0.6]{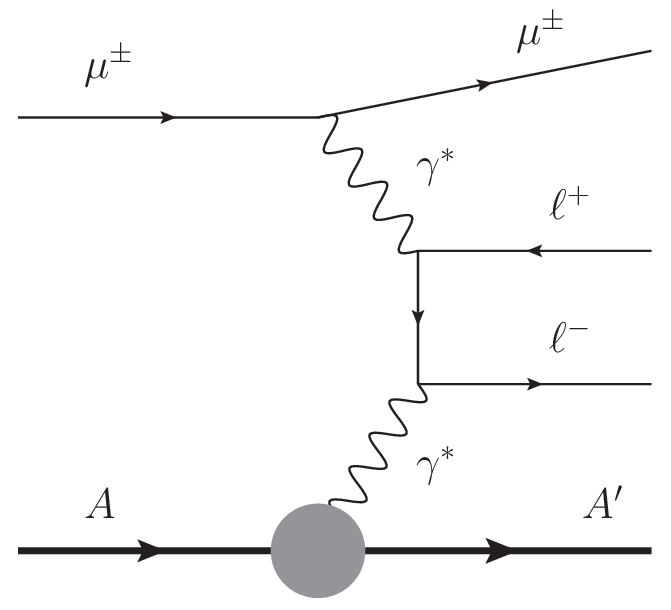} \;\;\;\;\;\;\;\;\;
    \includegraphics[scale=0.6]{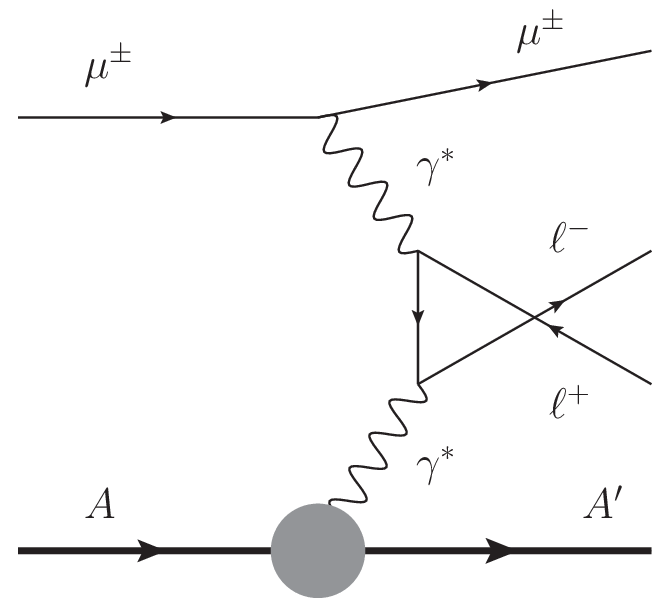} \\
    \includegraphics[scale=0.6]{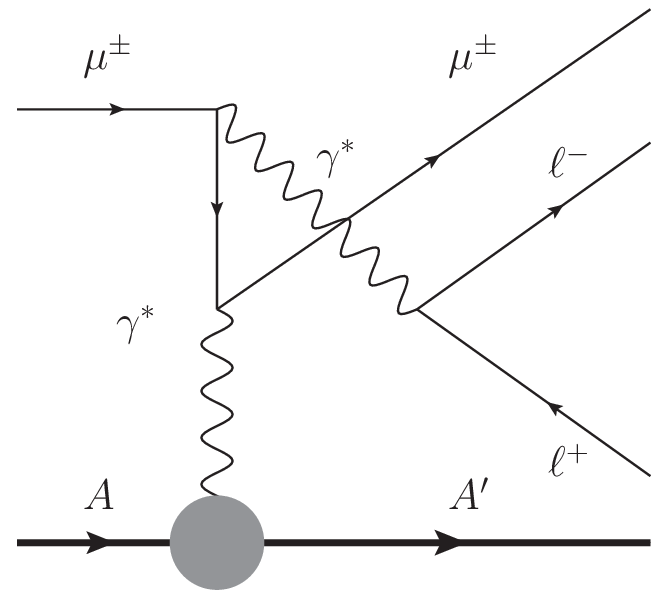} \;\;\;\;\;\;\;\;\;
    \includegraphics[scale=0.6]{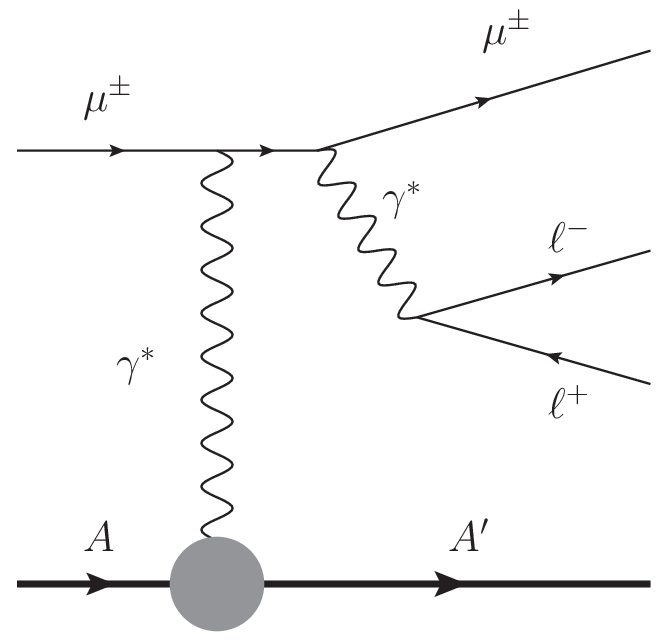}
    \end{tabular}
    \caption{Leading order diagrams for the electromagnetic production of a dilepton pair in a muon-ion scattering. The dilepton pair can be produced by the Bethe-Heitler process (upper diagrams) and by the muon bremsstrahlung channel (lower diagrams).}
    \label{fig:diagramas}
\end{figure}

\section{Formalism}
\label{sec:forma}
In this section, we will present a brief review of the formalism needed to describe the electromagnetic  production of a dilepton pair in a muon-ion scattering at the LHC energies. Such a process is represented by the reaction, 
\begin{eqnarray}
 \mu^{\pm}_{i} + A \rightarrow \mu^{\pm}_{f} + A + l^{+} + l^{-},
 \label{eq:reaction}
\end{eqnarray}
where $\mu^{\pm}_i$ ($\mu^{\pm}_f$) is the  muon in the initial (final) state and $l= e, \mu, \tau$ is the lepton  produced in the process. 
We will restrict our analysis to the coherent scattering, 
where the leptonic system scatters on the full nucleus and the nucleus remains intact in the final state. In this case, the cross-section is proportional to the square of the nuclear charge $Z^{2}$.   Our results will not take into account the contribution associated with the incoherent scattering  with the individual nucleons inside the nucleus, given that its contribution is proportional to $Z$, and therefore expected to be subleading. In principle, these two contributions can be experimentally separated, since the coherent scattering is expected to be characterized by a clean environment in the final state, i.e., without hadronic activity, in contrast with the incoherent case, where the nucleus break up. For a more detailed discussion about the incoherent contribution for the neutrino trident case, see e.g. Ref.~\cite{Francener:2024wul}.

At leading order, lepton pairs can be created electromagnetically by the Bethe-Heitler (photon-photon fusion) process and by bremsstrahlung from the interacting (muon and nucleus) particles. Previous studies have demonstrated that the contribution of nucleus bremsstrahlung is negligible \cite{Albright:1978mg,Ganapathi:1978qm,Kelner:1998mh}. As a consequence, in our calculations, we will only consider the contributions of the diagrams represented in Fig. \ref{fig:diagramas}. Because of charge-conjugation invariance, there is no interference between the Bethe-Heitler and bremsstrahlung diagrams in the total cross-section, but it is present in differential distributions. 
Experimentally, the muon trident process is characterized by a primary track associated with the incoming muon and three quasicollinear tracks for the outgoing leptons ($\mu_f^{\pm} l^+l^-$). Such a signature originates from the small deflection angle of the scattered muon, combined with the low momentum transfer of the muon to the nucleus. It is important to emphasize that a similar topology can also arise by the combination of the photon emission process ($\mu^{\pm}_{i} + A \rightarrow \mu^{\pm}_{f} + A + \gamma$) followed by gamma conversion ($\gamma \rightarrow l^+l^-$). 
Once in this process the photon has an interaction length of few millimeters in the matter, it would be possible to separate its contribution if we can reconstruct the interaction vertex properly.  However, such analysis is beyond the scope of the current work. An additional background for the muon trident process is the production of a pair of charged pions, with the pions being misidentified as muons. Since the hadronic interaction length  for pions with  the tungsten layers of FASER$\nu$(2) is of the order of $\lambda_{\mathrm{int}} \sim 10$ cm \cite{FASER:2020gpr}, the subtraction of this contribution is expected to be feasible.


Let us now focus on the muon trident process at the far-forward LHC detectors. In particular, we will consider the FASER$\nu$ and its proposed upgraded version,  FASER$\nu$2.  FASER$\nu$ is operating with 1.1 metric tons of tungsten distributed in 730 layers of 25 cm x 30 cm x 1.1 mm \cite{FASER:2018bac,FASER:2019dxq,FASER:2020gpr,FASER:2022hcn}. In our analysis, we will set the detector target length to be equal to 50 cm, since the muon identification, as well as the incoming and outgoing muon momentum measurements require a propagation over a few centimeters. We will assume an integrated luminosity of 250 fb$^{-1}$, which is expected for FASER$\nu$ data collection during run 3 of the LHC. It is important to emphasize that FASER$\nu$ detector is also expected to operate during Run 4 with a total integrated luminosity of 650 fb$^{-1}$. In contrast,  FASER$\nu$2 is expected to operate during the high luminosity LHC era in the Forward Physics Facility \cite{Anchordoqui:2021ghd,Feng:2022inv,FPF:2025bor}, being characterized by approximately 20 metric tons of tungsten and a time-integrated luminosity of 3 ab$^{-1}$. In what follows, we will discuss the muon trident production at FASER$\nu$, but the formalism can be directly extended for other far-forward LHC detectors.

The associated number of  muon trident events at FASER$\nu$ experiment can be expressed as \cite{Francener:2025pnr} 
\begin{eqnarray}
 N_{\text{events}} = \int^{1}_{0} \mathrm{d} x_{\mu^{\pm}} f(x_{\mu^{\pm}}) \sigma_{\mu^{\pm} + W} (E_{\mu^{\pm}_{i}}) \, ,
 \label{eq:events}
\end{eqnarray}
where $\sigma_{\mu^{\pm} + W} (E_{\mu^{\pm}})$ is the muon-tungsten cross-section for the process of interest, $f(x_{\mu^{\pm}})$ is the muon PDF, $E_{\mu^{\pm}_{i}}$ is the incident muon energy and $x_{\mu^{\pm}_{i}} = E_{\mu^{\pm}_{i}} / E_{p}$, with $E_p$ being the proton energy colliding at the ATLAS interaction point.
The muon PDF $f(x_{\mu^{\pm}})$  is the muon flux that reach the  detector and contains the information associated with detector geometry and its time exposure to collect data. For this quantity, we will use the muon and anti-muon fluxes in the forward direction of ATLAS simulated with FLUKA in Refs. \cite{Sabate-Gilarte:2023aeg,Battistoni:2015epi}. Recently,  Ref.~\cite{Francener:2025pnr} provided the muon flux in the LHAPDF format \cite{Buckley:2014ana}. As already pointed out in Ref.~\cite{Buckley:2014ana}, this simulation for the muon flux reaching far-forward LHC detectors can be accurately validated by measuring charged tracks in the spectrometer of the electronic detector and has effectively negligible uncertainties.

The muon-tungsten cross-section to the trident process, Eq.~(\ref{eq:reaction}), will be estimated using a modified version of the Monte Carlo event generator presented in \cite{Altmannshofer:2019zhy} for neutrino trident process. 
Such a generator was already adapted for a tungsten target in Ref.~\cite{Francener:2024wul}, where we also have  implemented a more precise description of the nuclear form factor, which was expressed in terms of the Fourier transform of the Woods-Saxon nuclear charge distribution \cite{Woods:1954zz}, with the parametrization provided in \cite{DeVries:1987atn}.
For the analysis performed in this paper, we have  generalized this improved version of Monte Carlo by the inclusion of  electromagnetic interactions initiated by a charged lepton. The cross-section is obtained taking into account all the lepton masses and the matrix element has been calculated with the package FeynCalc \cite{Mertig:1990an,Shtabovenko:2016sxi,Shtabovenko:2020gxv} without any approximation. In the case of a muon pair produced, there would be, in addition to the four Feynman diagrams of Fig.~\ref{fig:diagramas}, four more diagrams which consider the effects of two identical particles in the final state. Given that we are interested in a high energy incident muon ($>$ 10~GeV), the effects of Fermi-Dirac statistic in this kinematical range are negligible, as shown in Refs.~\cite{Kelner:1998mh,Russell:1971mf}. 

The above formalism can be directly extended for the calculation of the number of events associated with the production of QED bound states. In our analysis, we will focus on the production of singlet bound states, $(l^{+} l^{-})_S$, which can be produced by the fusion of two photons (For a more detailed discussion see, e.g., Refs. \cite{Ginzburg:1998df,Azevedo:2019hqp,Francener:2021wzx,Dai:2024imb,Bertulani:2023nch,Gninenko:2025hsv,Liang:2025sol,Francener:2024eep,Gargiulo:2023tci,Gargiulo:2024zyc,Gargiulo:2025pmu}). Using the equivalent photon approximation to describe the photon emission by the incoming muon \cite{epa}, the cross-section will be expressed by \cite{Dobrich:2015jyk}  
\begin{equation}
\begin{aligned}
    \sigma_{\mu^{\pm}_{i} + W \rightarrow \mu^{\pm}_{f} + W + (l^{+} l^{-})_S} =
    \frac{1}{2\pi E_{\mu^{\pm}_{i}}} \int   \mathrm{d}p_{t}^{2} \, \mathrm{d}\phi \, \mathrm{d}E_{(l^{+} l^{-})_S} \, \mathrm{d(cos}\, \theta )
     f_{\gamma/\mu} \left(\frac{E_{(l^{+} l^{-})_S}}{ E_{\mu^{\pm}_i}}, q_{t}^{2} \right) \frac{\mathrm{d}\sigma_{\gamma W}}{\mathrm{d(cos}\,\theta )}, 
\end{aligned}
\label{eq:muonium}
\end{equation}
with $p_{t}$ being the transverse momentum of the bound state, $\theta$ the angle between the produced state and the incident muon direction, and $\phi$ the angle between the transverse momenta of the  state and of the photon. The photon flux $f_{\gamma/\mu}(x, q_{t}^{2})$ is associated with the incident muon and can be written as \cite{epa,upc}
\begin{eqnarray}
     f_{\gamma/\mu}(x, q_{t}^{2}) = 
     \frac{\alpha}{2 \pi} 
     \frac{1+(1-x)^{2}}{x}
     \frac{q_{t}^{2}}{(q_{t}^{2}+ x^{2}q_{t}^{2})^{2}},
     \label{eq:EPA}
\end{eqnarray}
where $q_{t}$ is the transverse momentum transferred by the incident muon. The last ingredient in the Eq~({\ref{eq:muonium}}) is the photonuclear cross-section $\sigma_{\gamma W}$ that describes the Bethe-Heitler channel, which is evaluated with 
\begin{eqnarray}
    \frac{\mathrm{d}\sigma_{\gamma W}}{\mathrm{d(cos}\,\theta )} = 
    \frac{Z^{2}\alpha^{6} (-4 E_{(l^{+} l^{-})_S}^{2} t - 16m_l^{4}) }{4 m_l^{2} t^{2}} F(|t|)^{2} \, ,
    \label{eq:photonuclear}
\end{eqnarray}
where $F(|t|)^{2}$ is the nuclear form factor of the target nucleus of atomic number $Z$, which is a function of the squared momentum transfer by the nucleus, $t$, which can be approximated as 
\begin{eqnarray}
    t = 
    \frac{4m^{2}_l}{E_{(l^{+} l^{-})_S}^{2}} - p_{t}^{2}
    + 2 E_{(l^{+}  l^{-})_S} \, p_{t} \, \theta \, \mathrm{cos}\,\phi - E_{(l^{+} l^{-})_S}^{2}\, \theta^{2} \, .  
    \label{eq:t}
\end{eqnarray}
In what follows, we will estimate the cross-sections for the production of the true muonium  and positronium states in $\mu W$ interactions at FASER$\nu$ and FASER$\nu$2 detectors. Our main motivation is to verify if these detectors could be able to measure the  $(\mu^+ \mu^-)_S$ state, which was not yet observed.

\section{Results}
\label{sec:res}

In this section, we will present our predictions for the muon trident scattering with tungsten target at FASER$\nu$ and FASER$\nu$2 experiments. Initially, in order to check our implementation of the trident process,  we have estimated the cross-section for muon pair production in muon-carbon scattering for an incident muon of 160~GeV, which was previously calculated in Refs.~\cite{Abbiendi:2024swt,Lee:2023cwb}. Neglecting the nuclear form factor, as in Refs.~\cite{Abbiendi:2024swt,Lee:2023cwb}, we have obtained   195(1) nb for the corresponding cross-section, in agreement with the results of 196.3(9) nb and 196.74 nb derived using the  MESMER Monte Carlo \cite{Abbiendi:2024swt} and  an analytical expression presented in \cite{Lee:2023cwb}, respectively.

\begin{figure}[t]
	\centering
	\begin{tabular}{ccc}
    \includegraphics[scale=0.5]{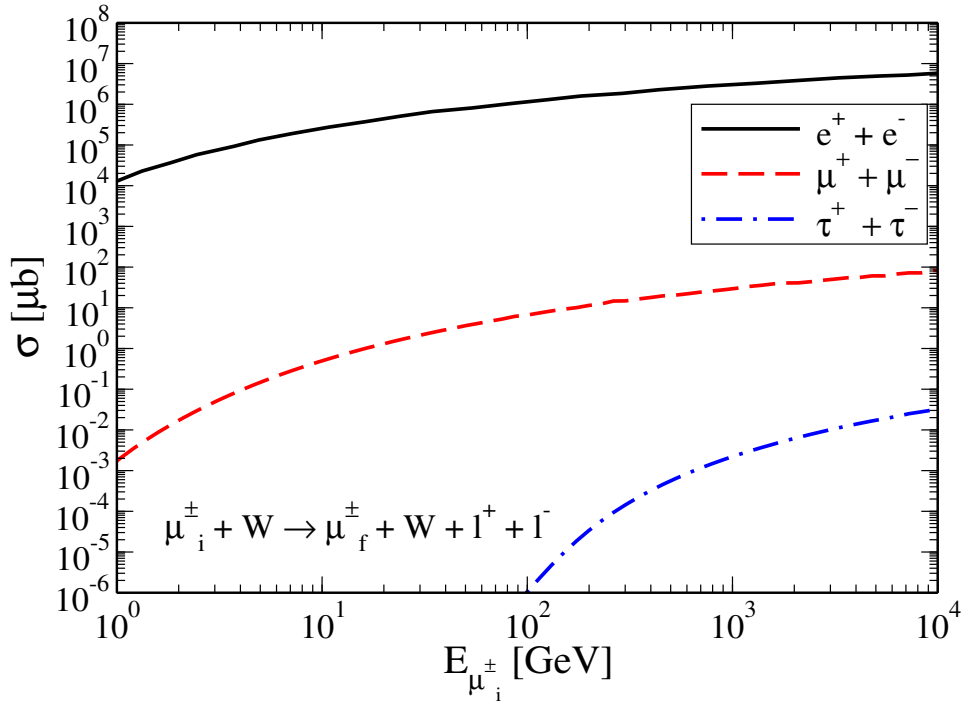}
    \end{tabular}
    \caption{ Total cross-section as a function of incident muon energy for distinct final states: pairs of electrons ( solid black line), muons (dashed red line) and taus (dash-dotted blue line).}
    \label{fig:sigmaTotal}
\end{figure}

In Fig.~\ref{fig:sigmaTotal} we present our results for the energy dependence of the total muon-tungsten cross-section in the regime covered by  far-forward LHC detectors. We present the cross-section for pair of electrons (solid black line), muons (dashed red line) and taus (dash-dotted blue line) produced in a coherent scattering, considering all the diagrams represented in Fig.~\ref{fig:diagramas}.  The cross-section for an electron-positron pair is approximately $10^{5}$ the process with a pair of muons produced, which is about $10^{3}$ times the process with a pair of taus produced by muons of TeV energy. For smaller energies of the incident muon, the cross-sections for taus in the final state decrease quickly, suppressed by the kinematical condition of $W_{\gamma^{*}\gamma^{*}} \ge 2 m_{\tau}$.

In what follows, using the total cross-section presented in the Fig.~\ref{fig:sigmaTotal} and the Eq.~(\ref{eq:events}), we will calculate the number of events expected at FASER$\nu$ detector during the run 3 considering a time integrated luminosity of 250~fb$^{-1}$. It is important to emphasize that the muon flux from \cite{Francener:2025pnr} does not take into account all the FASER$\nu$ length, but only 50~cm from the total of 80~cm. The first and last 15~cm of the target are supposed to be used to measure the ingoing and outgoing muon momentum through the multiple scattering in the tungsten. Therefore, the results for the event number are being underestimated by a factor 1.6 if we consider the real FASER$\nu$ length. Our results are obtained by summing over muon and anti-muon contributions and considering only incident (anti)muons with more than 10~GeV.

\begin{figure}[t!]
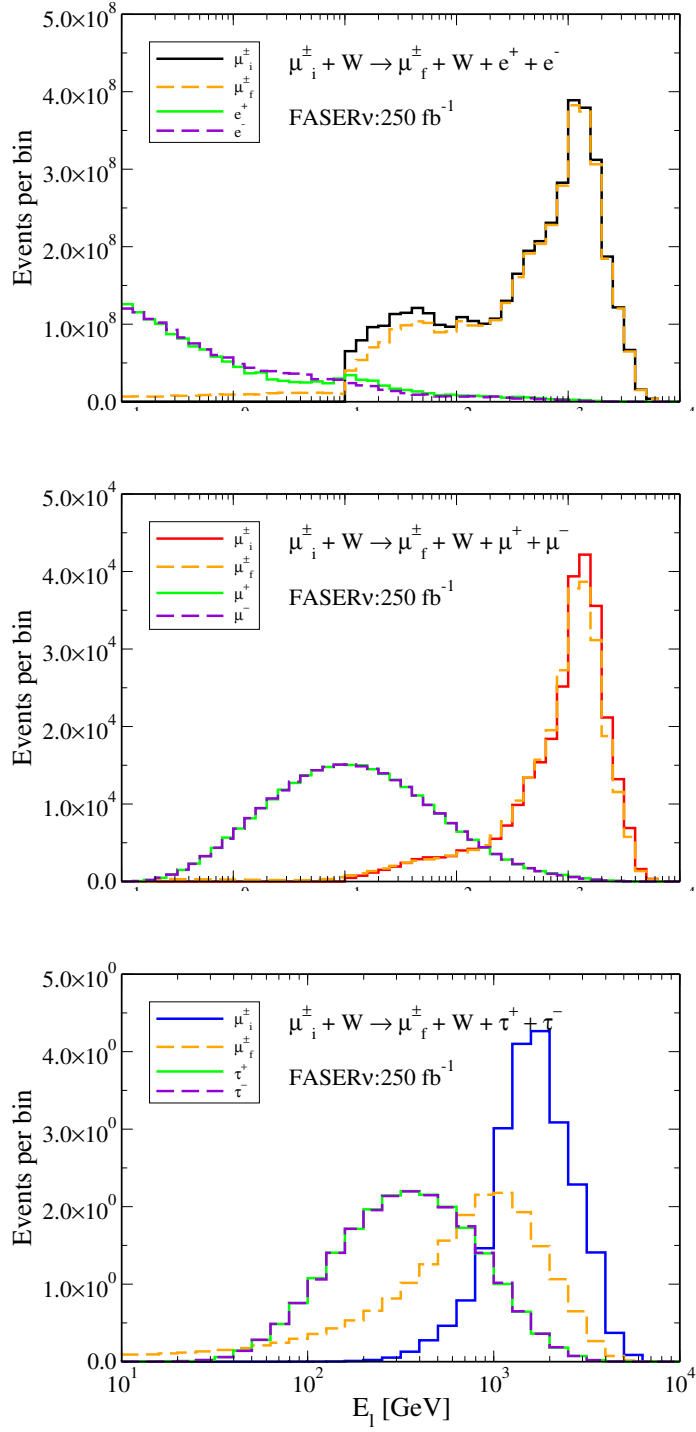

	\centering
	\begin{tabular}{ccc}
    \includegraphics[scale=0.34]{Figure/Events_E_ee_total.eps} \\
    \includegraphics[scale=0.34]{Figure/Events_E_mumu_total.eps} \\
    \includegraphics[scale=0.34]{Figure/Events_E_tautau_total.eps}
    \end{tabular}
    \caption{ Predictions for the expected number of events associated with the production of a pair of electrons (upper panel), muons (central panel) and taus (lower panel) by an incident muon, binned in the energy of each lepton. The results are for FASER$\nu$ during the run 3 of the LHC, considering an integrated luminosity of  250~fb$^{-1}$. }
    \label{fig:eventosBinEnergia}
\end{figure}

In Fig.~\ref{fig:eventosBinEnergia} we present our predictions for the number of muon trident events at FASER$\nu$ binned in the energy of the  leptons involved in the scattering, considering the production of a pair of electrons (upper panel), muons (central panel) and taus (lower panel).  We have that the expected number of events per bin is around $10^{9}$  for pairs of electrons, decreasing  by a factor of $10^{5}$ ($10^{9}$) for pairs of muons(taus). Our results indicate that the energies of the electron and positron produced in the interaction are small in the tungsten rest frame, being typically smaller than 1~GeV, with  the incident muon losing a small fraction of its energy. In contrast, $\mu^+$ and $ \mu^-$ leptons, produced by LHC muons, have energies of the order  $\mathcal{O}$(10~GeV). Finally, for the production of  $\tau^+ \tau^-$ pair, we have that the taus in the final state will have energy of  $\mathcal{O}$(300~GeV), and are produced by TeV energy muons.  As a consequence, in this case, the incident muon loses a large fraction of its energy. At lower energies, the number of events is strongly suppressed due to the behavior of the cross-section in this kinematical range (See Fig. \ref{fig:sigmaTotal}). It is important to emphasize that a tau with an energy in the range of 100~GeV$-$1~TeV energy has decay length of the order of 5~mm$-$5~cm, being possible to reconstruct its decay with the emulsion detector between the tungsten plates present at FASER$\nu$ detector.

\begin{figure}[t]
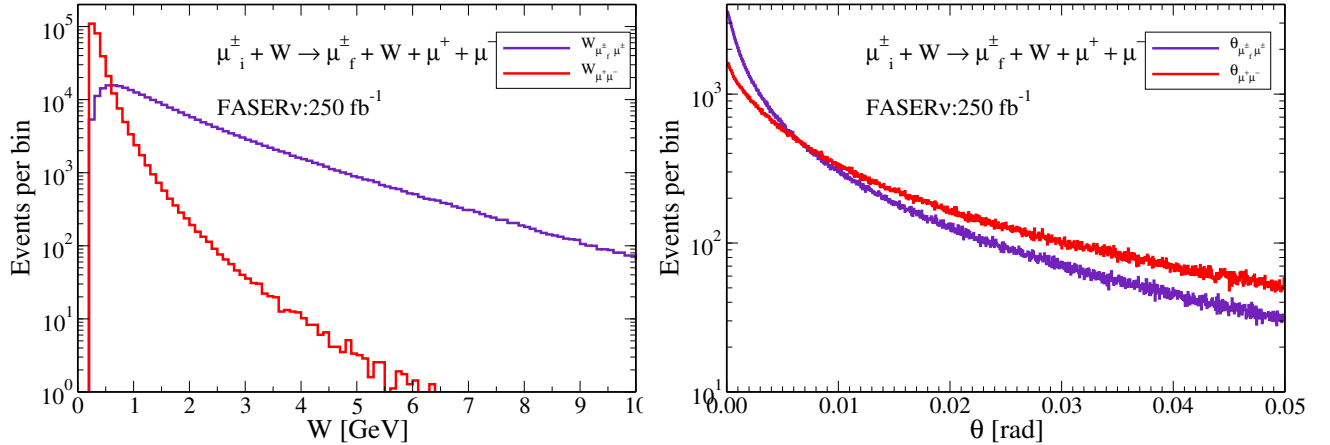

	\centering
	\begin{tabular}{ccc}
    \includegraphics[scale=0.34]{Figure/Events_W_mumu_total.eps} 
    \includegraphics[scale=0.34]{Figure/Events_theta_mumu_total.eps}
    \end{tabular}
    \caption{ Predictions for the number of events associated with the electromagnetic $\mu^+\mu^-$ production by an incident muon,  binned in the invariant mass (left panel) and angular opening (right panel) of two muons in the final state. The results are for FASER$\nu$ during the run 3 of the LHC considering an integrated luminosity of  250~fb$^{-1}$. }
    \label{fig:events_W_theta}
\end{figure}

The electromagnetic $\mu^+\mu^-$ production by an incoming muon is characterized by two identical particles in the final state. As the current detectors are not able to identify what muon is produced in the vertex of the incident muon, it is interesting to analyze if it is possible to distinguish  the produced pair ($\mu^+\mu^-$) from a pair associated with a produced muon + outgoing muon ($\mu^{\mp}\mu_f^\pm$). In the Fig.~\ref{fig:events_W_theta} we show our results for the muon trident binned in the invariant mass (left panel) and opening angle (right panel) for these systems of  muon pairs. We have considered a bin of 0.1~GeV for the invariant mass and 0.05~mrad for the opening angle, which is the angular sensitivity of FASER$\nu$ detector \cite{FASER:2025qaf}. 
The  produced pair has a lower invariant mass than the $\mu^{\mp}\mu_f^\pm$ pair, which is associated with the higher energy of the muon in the final state that was produced in the incident muon vertex. The produced pair also has in general a larger opening angle than the $\mu^{\mp}\mu_f^\pm$ one, given the small deflection of the incident muon after the scattering.

\begin{center}
	\begin{table}[t]
		\begin{tabular}{|c|c|c|c|c|}
			\hline
			\hline 
Final state		      & FASER$\nu$           & FASER$\nu$ ($E_l \ge 10$~GeV)& FASER$\nu$2 & FASER$\nu$2 ($E_l \ge 10$~GeV) \\
			\hline 	
			\hline
$e^{+} + e^{-}$ 	  & 4.10$\times 10^{10}$ & 7.23$\times 10^{8}$ & 6.10$\times 10^{12}$ & 1.08$\times 10^{10}$
\tabularnewline
$\mu^{+} + \mu^{-}$   & 2.61$\times 10^{5}$  & 1.03$\times 10^{5}$  & 3.87$\times 10^{7}$ & 1.53$\times 10^{7}$ 
\tabularnewline
$\tau^{+} + \tau^{-}$ & 21.83                & 19.97                & 3.25$\times 10^{3}$ & 2.97$\times 10^{3}$ 
\tabularnewline
			\hline 
            \hline
		\end{tabular}
		\caption{ Number of events associated with the electromagnetic dilepton production in $\mu W$ interactions at FASER$\nu$ and FASER$\nu$2 detectors, derived assuming integrated luminosities of $\mathcal{L}_{\rm pp}=250$ fb$^{-1}$ and 3 ab$^{-1}$, respectively. Results derived with and without the inclusion of an energy cut on the final-state leptons. }
		\label{table:Nevents_open}
	\end{table}
\end{center}

In the Table~\ref{table:Nevents_open} we present the number of events expected at FASER$\nu$ and FASER$\nu$2 during the run 3 and the high luminosity era of the LHC, derived considering an integrated luminosity of 250~fb$^{-1}$ and 3~ab$^{-1}$, respectively. The impact of an energy cut on the final-state leptons is also estimated. For the energy cut, we are assuming leptons in the final state with at least 10~GeV each. The total of events expected at FASER$\nu$ detector for pairs of electrons, muons and taus are larger than $10^{10}$, $10^{5}$ and 10, respectively. When  the energy cut on the final-state leptons is considered, the number associated with pairs of electrons, muons and taus decrease by factors of $56.7 $, $2.5 $ and $1.1 $, respectively. The events at FASER$\nu$2 at HL-LHC increase by a factor of $\approx 150$. Such results  indicate that FASER$\nu$ will be able to perform a detailed study of the electromagnetic production of $e^+e^-$ and $\mu^+\mu^-$ pairs. In addition, the first observation of the  $\tau^+ \tau^-$ pair production in the trident muon process is, in principle, feasible using this detector.



\begin{figure}[t]
	\centering
	\begin{tabular}{ccc}
    \includegraphics[scale=0.33]{Figure/cross_section_muon_trident_muonium.eps} 
    \includegraphics[scale=0.34]{Figure/TrueMuonium_dist_W_100GeV.eps}
    \end{tabular}
    \caption{ Total cross-sections for true muonium and open muon pair production in $\mu W$ interactions as a function of incident muon energy (left panel). Differential cross-section as a function of the invariant mass of the true muonium and open muon pair, produced in $\mu W$ interactions for an incident muon of 100~GeV (right panel). 
    }
    \label{fig:sigmaMuonium}
\end{figure}

Finally, let us consider the  electromagnetic production of a true muonium $(\mu^+ \mu^-)_S$ by an incoming muon and estimate the associated cross-section and expected number of events.  
In Fig.~\ref{fig:sigmaMuonium} (left panel) 
we present the total cross-section as a function of incident muon energy. For comparison, the results for the production of open $\mu^+\mu^-$ pair is also presented. Our results indicate that the true muonium cross-section is smaller than the open pair production by a factor of $10^{4}$ and $10^{6}$ for an incident muon of 1~GeV and 10~TeV, respectively. The invariant mass distribution of the cross-section for a pair of muons produced by an incident muon of 100~GeV is presented in the right panel of Fig.~\ref{fig:sigmaMuonium}. While the open pair of muons has a continuum spectrum starting in the production threshold, the bound state is characterized by a resonance peak at  $ W \approx 2m_\mu$ with a width of the order of $\alpha^5 m_\mu / 2$.
In FASER-like detectors, the signal of a true muonium is expected to be a pair of muons with approximately the same energy and highly collimated, given that the true muonium is expected to dissociate in the tungsten plates after propagates less than 0.1~mm. 
The associated number of events expected at FASER$\nu$ and FASER$\nu$2 during the run 3 and the high luminosity era of the LHC, considering an integrated luminosity of 250~fb$^{-1}$ and 3~ab$^{-1}$, respectively, are presented  in the Table~\ref{table:Nevents}.  Our estimation for total true muonium events at FASER$\nu$ is less than one, but around 60 events at FASER$\nu$2. In contrast, for the positronium production, we predict that the number of events will be larger than $10^5$ ($10^7$) at  FASER$\nu$ (FASER$\nu$2). When we assume a cut on the final-state lepton energies of 10~GeV, the number of events decreases by a factor of $\approx 1.9$ and $\approx 16.2$ for true muonium and positronium production, respectively.

\begin{center}
	\begin{table}[t]
		\begin{tabular}{|c|c|c|c|c|}
			\hline
			\hline 
Final state		        & FASER$\nu$          & FASER$\nu$ ($E_l \ge 10$~GeV) & FASER$\nu$2         & FASER$\nu$2($E_l \ge 10$~GeV) \\
			\hline 	
			\hline
$(e^{+} e^{-})_{S}$     & 1.51$\times 10^{5}$ & 9.30$\times 10^{3}$ & 2.24$\times 10^{7}$ & 1.38$\times 10^{6}$
\tabularnewline
$(\mu^{+} \mu^{-})_{S}$ & 0.38                & 0.20                & 57.04               & 29.76
\tabularnewline
			\hline 
            \hline
		\end{tabular}
		\caption{Number of events associated with the electromagnetic  production of  QED bound states in $\mu W$ interactions at FASER$\nu$ and FASER$\nu$2 detectors, derived assuming integrated luminosities of $\mathcal{L}_{\rm pp}=250$ fb$^{-1}$ and 3 ab$^{-1}$, respectively. Results derived with and without the inclusion of an energy cuts on the final-state leptons.}
		\label{table:Nevents}
	\end{table}
\end{center}

\section{Summary}
\label{sec:sum}

In this paper, we have investigated the electromagnetic production of lepton pairs  and QED bound states in muon-tungsten interactions at far-forward LHC detectors. In particular,  we have estimated the  cross-sections for the dilepton production considering the contributions of the  Bethe-Heitler and bremsstrahlung processes in the coherent $\mu W$ scattering,  and  derived the predictions for the event rates and associated distributions at FASER$\nu$ detector. Our results indicated that this far-forward detector will be able to perform a detailed study of the electromagnetic $e^+ e^-$ and $\mu^+ \mu^-$ production and  constrain the corresponding cross -sections. In contrast, for the $\tau^+ \tau^-$ production, we predict that around twenty pairs of taus will be produced in this detector, which indicates that such process can, in principle, be observed for the first time even during the run 3. In addition, we predict that the event rates will increase by two orders of magnitude at FASER$\nu$2. Finally, we have also estimated the cross-sections and event rates for the production of QED bound states. Our results indicated that a large number of positronium will be produced in the far-forward LHC detectors. On the other hand, the measurement of a true muonium  is, in principle, only feasible at FASER$\nu$2 detector. Such results motivate a more detailed analysis of the experimental separation of these states, which we postpone for a future publication.

\begin{acknowledgments}
 The authors would like to thank Felix Kling for helpful comments and discussions. R.F. acknowledges support from the Conselho Nacional de Desenvolvimento Cient\'{\i}fico e Tecnol\'ogico (CNPq, Brazil), Grant No. 161770/2022-3. V.P.G. was partially supported by CNPq, FAPERGS and INCT-FNA (Process No. 464898/2014-5). G.R.S. is supported by the CAPES doctoral
fellowship 88887.005836/2024-00.
\end{acknowledgments}

\hspace{1.0cm}

\end{document}